\documentclass[11pt]{article} 

\usepackage{dsfont}

\usepackage{comment}

\usepackage{amsmath}

\usepackage{epsfig}
\usepackage{color}
\usepackage{float}
\usepackage{graphicx}
\usepackage{url}
\usepackage{natbib}
\usepackage{epstopdf}
\usepackage[font=small, labelfont=bf, margin=0.50 in]{caption}
\usepackage{multirow}
\usepackage{subfloat}
\usepackage{subcaption}
\usepackage{indentfirst}
\usepackage[letterpaper, left=0.5 in, right=0.5 in, top=0.75 in, bottom=0.75 in]{geometry}
\usepackage[symbol]{footmisc}
\usepackage{pdfpages}
\usepackage [autostyle, english = american]{csquotes}

\MakeOuterQuote{"}

\def\correspondingauthor{\footnote{Correspondence:  salt@yorku.ca\vspace{2mm}}}

\graphicspath{ {images/} }

\begin{document}

\title{Optimal Allocation to Deferred Income Annuities}
\author{F. Habib$^{\dagger}$, H. Huang$^{\dagger}$, A. Mauskopf$^{\dagger}$, B. Nikolic$^{\dagger}$, T.S. Salisbury\correspondingauthor $^{*}$\footnote{\hangafter=1\hangindent=1.7em  Faisal Habib and Branislav Nikolic are instructors at the Schulich School of Business and researchers at CANNEX Financial Exchanges Limited; Adam Mauskopf is a graduate student at York University; Huaxiong Huang and Thomas S. Salisbury are faculty in the Department of Mathematics \& Statistics at York University, all in Toronto, Canada.
\vspace{2mm}}}

\maketitle

\setlength{\parindent}{0pt}
\renewcommand*{\thefootnote}{\arabic{footnote}}

\abstract{\noindent In this paper we employ a lifecycle model that uses utility of consumption and bequest to determine an optimal Deferred Income Annuity (DIA) purchase policy.  We lay out a mathematical framework to formalize the optimization process. The method and implementation of the optimization is explained, and the results are then analyzed. We extend our model to control for asset allocation and show how the purchase policy changes when one is allowed to vary asset allocation. Our results indicate that (i.) refundable DIAs are less appealing than non-refundable DIAs because of the loss of mortality credits; (ii.) the DIA allocation region is larger under the fixed asset allocation strategy due to it becoming a proxy for fixed-income allocation; and (iii.) when the investor is allowed to change asset-allocation, DIA allocation becomes less appealing.  However, a case for higher DIA allocation can be made for those individuals who perceive their longevity to be higher than the population.}

\setlength{\parindent}{0pt}

\section{Introduction}
Over the last few decades, we have seen a shift away from \emph{Defined Benefit} (DB) pension plans to \emph{Defined Contribution} (DC) pension plans (\cite{ebri-dbdc-2014}).  This shift is usually attributed to the fact that many firms have not been able to fully fund the future liabilities of a DB plan, for a myriad of reasons -- high among them being the improvement in human longevity.  In 2014, only 2\% of private sector workers participated in a DB pension plan.  In a DB plan, future income is guaranteed by the employer, whereas no such guarantee is made in a DC plan.  The employer simply contributes to the plan on behalf of the employee (and often times matches the employee contributions).

\medbreak
In the United States, prior to 2014, holding an income annuity within a retirement plan was not considered a tax effective strategy because the rules for retirement accounts required that minimum distributions began at age $70\frac{1}{2}$~.  In 2014, the Internal Revenue Service (IRS) revised its rules with respect to Required Minimum Distributions (RMD) and now with a Qualified Longevity Annuity Contract (QLAC), a plan participant may fund income annuities with assets from employer sponsored qualified plans (i.e. 401k plans), and be exempt from RMD until the start of payments.   Insurance companies now offer \emph{Deferred Income Annuities} (DIA) which are retirement income products that allow one to purchase an annuity prior to retirement, with the start of income deferred to a future date. This then provides an opportunity to the plan participant to "pensionize\textsuperscript{\textregistered}"\footnote{Pensionize is a registered trademark of CANNEX Financial Exchanges Limited.} some of her retirement assets thus partially replicating the benefits of a DB plan.

\medbreak
In this paper, we are interested in questions such as: should one purchase DIAs now, or wait until retirement to invest in a {\it Single Premium Immediate Annuity} (SPIA)? Should one purchase DIAs early in a lump sum up front, or should the purchase strategy evolve over time (eg. dollar cost average into an annuity)?  We are motivated by \cite{MY2007} in which the authors examined the optimal annuitization (using a SPIA), investment and consumption strategies of a utility maximizing agent.  In this paper, we start with a similar approach but in the context of a DIA.  Later we expand our model to control for asset allocation.

\subsection{Literature review}
\cite{yaari65} demonstrated that in a lifecycle model with no bequest motives all economic agents should hold actuarial notes as opposed to liquid assets.  The rationale is that the returns from annuities will dominate all other assets because the "dead" subsidize the "living".  The reality, however, is very different as few individuals consciously choose to annuitize all of their assets.  This lack of annuitization has become what is known as the \emph{annuity puzzle} and has led to an extensive body of research on the optimal purchase of income annuities within the context of portfolio choice -- see for example, \cite{campbell-viceira-2002}, \cite{neuberger-2003}, \citet{cocco-gomes-maenhout-2005}, \cite{horneff-maurer-mitchell-stamos-2009}, \cite{koijen-nijman-werker-2011}, and \cite{hainuat-deelstra-2014}.

\medbreak
\cite{MY2007} examine the optimal annuitization, investment and consumption strategies of a utility maximizing agent and conclude that that there is an incentive to delay annuitization in cases where annuitization has to take place in an "all or nothing" format and that this incentive occurs even in the absence of bequest motives.  \cite{blanchett-2015} utilizes a model to purchase annuities at anytime -- immediate annuitization upon retirement and dynamic annuitization (how much additional to buy while in retirement) based on a funding ratio to minimize portfolio shortfall risk.  \cite{milevsky-moore-young-2006} employ a model to minimize the probability of financial ruin in determining annuity purchase strategies.

\medbreak
Much of this research involves the decision around a SPIA purchase either at the beginning of retirement or during the time spent in retirement.   \cite{huang-milevsky-young-2017} consider the optimal purchasing of DIA. They do so in the context of a consumer wishing to maximize the expected utility of her annuity income at a fixed time in the future. In their model, complete annuitization by this fixed time is assumed, and they study the impact of a fluctuating annuity payout yield on the DIA purchase. They find that DIA purchase will be gradual, and will occur only when the payout yield rises above some level. 

\medbreak
We build upon the latter work, and attempt to understand in more detail the annuity purchase decision for an individual who seeks to maximize her expected utility within the context of employer sponsored retirement plans. First of all, we do not assume a cash lump sum that must all get annuitized at or before a fixed time horizon. Instead we allow investment alternatives to the annuity, so deciding how much to annuitize becomes part of our problem. Our model is split into two phases -- pre-retirement (in which DIAs are purchased over time) and post-retirement, in which the retiree consumes optimally from DIA income, investment accounts, and (exogenous) pensions. Our approach in post-retirement employs the same framework as \cite{vulcan2011} and \cite{habib-huang-milevsky-2017}.  In post-retirement, we solve the \cite{merton71} problem in which investment returns as well as human lifetimes are random but one in which pension annuities are available.  The annuity purchase decision (in our case, purchases of DIA) happens only in pre-retirement. Our annuity payout factor is a function of age, but in future work we hope to incorporate interest rate uncertainty, which would lead to the kinds of payout factor volatility which is the focus of \cite{huang-milevsky-young-2017}. 

\medbreak
Our conclusion is that DIA purchase can occur either gradually over time, or in a lump sum (either at initiation or at retirement), or both, depending on what investment alternatives are modelled. For a given level of existing DIA there will be a wealth cutoff that varies with age, below which one does not purchase additional DIA, but above which one does. We pay particular attention to the degree to which DIA purchase is influenced by asset allocation (in the investment account) and mortality credits (ie. the way DIA refunds are structured in case of pre-retirement death).

\subsection{Agenda of the paper}
In section \ref{section:IncomeAnnuities} we provide a brief introduction to income annuities and the model for human mortality.  For a more extensive development and explanation, see \cite{cri2006}.  In section \ref{section:WealthDynamics} we describe the various account dynamics and in section \ref{section:HJB} we discuss the formulation of our mathematical problem, the resulting PDE, and its boundary conditions.  In section \ref{section:Observations1} we present and discuss our results.  In section \ref{section:AssetAllocation}, we reformulate the original problem to allow for asset allocation and present our results in that case in section \ref{section:Observations2}.

\section{Income annuities} \label{section:IncomeAnnuities}
An annuity is an insurance product that pays an annuitant (purchaser) an income for the rest of her life.   The stochastic present value of a pension annuity that pays \$1 is

\begin{equation}
a_x = \int_{0}^{T_x} e^{-rs} ds = \int_{0}^{\infty} e^{-rs}  \mathds{1}_{\{T_x \geq s\}} ds
\end{equation}

where $T_x$ is the remaining lifetime random variable.  The expected value of the the above random variable is known as the \emph{Immediate Pension Annuity Factor (IPAF)}

\begin{equation}
\bar{a}_x = \mathbf{E}\left[ \int_{0}^{\infty} e^{-rs} \mathds{1}_{\{T_x \geq s\}} ds \right] = \int_{0}^{\infty} e^{-rs} {} _{s}{p}_{x} ds
\end{equation}

where $_{t}{p}_{x}$ is the $t$-year probability of survival of an $x$-year old individual.\\

A Deferred Income Annuity (DIA) is similar to an immediate income annuity, however, it starts paying income after a deferral period $\tau$.  The expected value of a DIA that pays \$1 after the deferral period is

\begin{equation}
{}_{\tau}\bar{a}_{x} =  \int_{\tau}^{\infty} e^{-rs} {}_{s}{p}_{x} ds = e^{-\tau r} {}_{\tau}p_{x}  \int_{0}^{\infty} e^{-rs} {}_{s}p_{(x+\tau)} ds 
\end{equation}
which can be simplified as
\begin{equation}
\label{abarnoref}
{}_{\tau}\bar{a}_{x}  = e^{-\tau r} {}_{\tau}p_{x} \bar{a}_{(x+\tau)}.
\end{equation}

The above equation is for a DIA which does not offer any refund if the investor dies before retirement. In reality, most DIA's currently available in the market offer a refund, so we want to model the refund aspect of the DIA as well. Consider the following version of a DIA:  each unit of DIA that the investor buys is refunded to her should she die, at the value that it is worth on the day of death.  For example, if the investor buys two units of DIA (that is to say, she is now entitled to \$2 per year in retirement) at \$100, and then dies in one year when the DIA is worth \$110, then her estate would receive \$220 . 

\medbreak
With the above refund scheme, there is essentially no financial impact on legacy due to pre-retirement death, since the entire sum of money deposited into the account is returned, with interest. This theoretical DIA will be priced at 
   \begin{equation}
   \label{a_xnorisk}
   _\tau\bar{a}_x = \bar{a}_{(x+\tau)}e^{-r\tau}.
   \end{equation}

Equation (\ref{a_xnorisk}) is almost identical to the equation (\ref{abarnoref}); we simply remove the factor $_{\tau}p_x$ since that is the probability the refund will not take place.

\medbreak
We will analyze DIAs that interpolate between these extremes.  We use a weighted average of the no-refund DIA, and the full-refund DIA, and introduce the parameter $Q$ that takes values between 0 and 1. That is,
   \begin{equation}
   \label{a_xhaircut}
   {_\tau\tilde{a}_x}= {\bar{a}_{x+\tau}}e^{-r\tau}[_{\tau}p_x (1-Q) +Q].
   \end{equation}
   
\medbreak
As we shift $Q$ from 0 to 1, we change the nature of the DIA. A $Q$ value of 0 implies a DIA with no refund (i.e. it includes mortality credits), while a $Q$ value of 1 implies a DIA with a full refund (i.e. it includes no mortality credits).  In case a refund is issued, the value that gets returned is only the portion $Q$ of the full refund DIA.  That is to say, the refund, $K_t$ is defined as
    \begin{equation}
   \label{refundhaircut}
   K_t= {\bar{a}_{x+\tau}}e^{-r(\tau-t)} Q.
   \end{equation}

\medbreak
  We use the Gompertz-Makeham Law of Mortality to model human mortality in which, $\lambda(x)$,  the \emph{Instantaneous Force of Mortality} (IFM) is defined as

\begin{equation}
\lambda(x) = \lambda_0 + \frac{1}{b}e^{(x-m)/b}
\end{equation}
 where \emph{x} is the age, \emph{$\lambda_0$} captures the death rate attributable to accidents, \emph{m} is the modal value of life\footnote{strictly speaking, it is the mode only when $\lambda_0=0$}, and \emph{b} is the dispersion coefficient.  The conditional survival probability is then
\begin{equation}
 {{_t}{p_x}} = e^{-\lambda_0t + \left(1-e^\frac{t}{b}\right)e^{\frac{x-m}{b}} }
\end{equation}
  where once again \emph{x} is the age of the individual and \emph{t} is the number of survival years.  This analytical representation of survival probabilities can be calibrated to a mortality table that is readily available from the Society of Actuaries.
\medbreak



\section{Wealth dynamics} \label{section:WealthDynamics}
When it comes to retirement savings accounts, a DC plan (or a 401k plan) allows the plan participant to make a contribution (typically matched by the employer up to a certain limit) that is invested into a portfolio of funds (mutual funds, ETFs, or a portfolio of stocks and bonds) as per the instructions of the plan participant.   We consider an account whereby the plan participant can not only invest in a portfolio of stocks and bonds, but also be able to purchase DIAs periodically.  We assume that the account value follows a Geometric Brownian Motion (GBM) and model the wealth dynamics in two phases -- \emph{pre-retirement} and \emph{post-retirement}.  In the pre-retirement phase, the client is allowed to make deposits and purchase DIAs.  In the post-retirement phase, the DIA generates income and the client (now retired) spends optimally from the remaining investment account.

\medbreak
Before a fixed retirement date $\tau$, i.e., $t<\tau$, the wealth dynamics for the retirement account $W_t$ are given by
\begin{equation}
dW_t=(\mu W_t+\nu-g_t)dt+\sigma W_tdB_t
\end{equation}
where $\mu$ and $\sigma$ are the drift and volatility of the investment portfolio, $\nu$ is a steady exogenous deposit stream to the retirement funds, and $g_t\geq 0$ is the rate for purchasing deferred annuities (DIA) with income starting at time $\tau$.
Of course, we require that $g_t\le\nu$ when $W_t=0$, in order to ensure the existence of a non-negative solution $W_t$.

\medbreak
The income from the accumulation of DIA, denoted by $I_t$, is given by
\begin{equation}
dI_t=\frac{g_t}{{}_{(\tau-t)}\tilde{a}_{(x+t)} } dt.
\end{equation}

In post-retirement, $t\geq \tau$, the wealth dynamics are given by
\begin{equation}
dW_t=(\mu W_t+I_t+\pi-c_t)dt+\sigma W_tdB_t
\end{equation}
where $\pi$ is the exogenous pension income rate (sourced from government pensions such as social security benefits), $I_t$ is the aggregate income from the purchased DIAs, and $c_t$ is the consumption rate.

\section{Optimal allocation to DIAs} \label{section:HJB}
The basic question in pre-retirement ($t<\tau$) is to determine the optimal choice for $g_t$, while in post-retirement ($t\geq \tau$) we want to know the optimal $c_t$. To do that we define a value function (one that also incorporates a bequest motive) with  $g_t$ and $c_t$ as our control variables
\begin{equation}
J(t,w,I) = \max_{g_s,c_s}E\left[\int^\infty_t e^{-\rho s} \left._sp_x(U(c_s)+\lambda_{x+s} \left\lbrace  U(W_s+\pi) \mathds{1}_{t\geq\tau} \right. +  U(W_s+K_s+\nu) \mathds{1}_{t<\tau}  \right\rbrace ) \right|W_t=w,I_t=I\bigg]. \nonumber
\end{equation}

Here  $U(x)$ is a utility function. We typically adopt the Constant Relativity Risk Aversion (CRRA) utility model,  with $U(x) = \frac{x^{1-\gamma}}{1-\gamma}$, where $\gamma$ is the subjective risk aversion of a particular investor. 
Note that upon death (say at time $s$), prior to starting income, a cash refund of $K_s$ is added to the total (estate) wealth $W_s$. We also add $\pi$ (resp. $\nu$) to the estate, upon post-retirement (resp. pre-retirement) death. Mathematically this provides computational stability, because it forces all the terms of the PDE to remain bounded as $w\downarrow 0$. Economically it reflects the reality that cash flows and income streams do not stop abruptly with death, but continue for some time afterwards (taken here to be a year). 

\subsection{Hamilton-Jacobi-Bellman equation}
For any admissible choices of $g_t\geq 0$ and $c_t\geq 0$ the value function satisfies the Hamilton-Jacobi-Bellman (HJB) equation 
\begin{eqnarray}
0&\geq& J_t+(\mu w+\nu-g_t)J_w+\frac{1}{2}(\sigma w)^2J_{ww} \label{pre}\\
&&+\frac{g_t}{{}_{(\tau-t)}\tilde{a}_{(x+t)} }J_I+\lambda_{x+t} U(w+K+\nu)-(\rho+\lambda_{x+t})J \nonumber
\end{eqnarray}
for $t<\tau$, i.e. \emph{pre-retirement}, and
\begin{eqnarray}
0&\geq& J_t+(\mu w+I_t+\pi-c_t)J_w+\frac{1}{2}(\sigma w)^2J_{ww} \label{post}\\
&&+\lambda_{x+t} U(w+\pi)+U(c_t)-(\rho+\lambda_{x+t})J \nonumber
\end{eqnarray}
for $t\geq\tau$, i.e. \emph{post-retirement}. The inequalities become equalities when the optimal choices of $g_t$ and $c_t$ (denoted by $g^*_t$ and $c^*_t$) are used.

\medbreak
We apply the first order condition to obtain the optimal consumption rate $c^*_t$, which is given implicitly by
\begin{equation}
\label{firststep}
-J_w+U'(c^*_t)=0
\end{equation}
for any concave utility function $U(x)$. The HJB equation becomes
\begin{eqnarray}
0&=& J_t+(\mu w+I_t+\pi-c^*_t)J_w+\frac{1}{2}(\sigma w)^2J_{ww} \label{post2}\\
&&+\lambda_{x+t} U(w+\pi)+U(c^*_t)-(\rho+\lambda_{x+t})J \nonumber
\end{eqnarray}
for $t\geq \tau$.

\medbreak
For the pre-retirement stage, the problem is different as the functional is linear in $g_t$. It is instructive to examine the variation
\begin{equation}
\delta \mathcal{H}:=\mathcal{H}'\delta g_t 
\end{equation}
where
\begin{equation}
\mathcal{H}:=-g_tJ_w+\frac{g_t}{{}_{(\tau-t)}\tilde{a}_{(x+t)} }J_I, \;\; \mathcal{H}'=\frac{J_I}{{}_{(\tau-t)}\tilde{a}_{(x+t)} }-J_w .
\end{equation}
Since $\delta g_t$ is non-negative, the sign of $\mathcal{H}'$ determines optimality.

\medbreak
If $\mathcal{H}'>0$, it is optimal to annuitize, since that makes $\delta \mathcal{H}>0$. On the other hand, if $\mathcal{H}'<0$, it is optimal not to annuitize ($g_t=0$) since annuitization makes $\delta \mathcal{H}<0$, which decreases the value function. The annuitization boundary is therefore given by $\mathcal{H}'=0$, or
\begin{equation}
\frac{J_I}{{}_{(\tau-t)}\tilde{a}_{(x+t)} }=J_w. \label{boundary}
\end{equation}

In addition, when it is optimal not to annuitize, equality holds in (\ref{pre}). Otherwise when it is optimal to annuitize, including on the annuitization boundary (\ref{boundary}), the inequality in (\ref{pre}) holds. Therefore, we can write down a single equation for all the cases considered here
\begin{eqnarray}
\label{laststep}
&& \max \bigg \{J_t+(\mu w+\nu)J_w+\frac{1}{2}(\sigma w)^2J_{ww} +\lambda_{x+t} U(w+K_t+\nu)-(\rho+\lambda_{x+t})J,  \nonumber \\
&& \quad\quad\quad   J_I-{}_{(\tau-t)}\tilde{a}_{(x+t)} J_w\bigg\} =0 \label{pre2}
\end{eqnarray} 
for $t<\tau$.

\medbreak
The solution of the above equation provides the boundary between the annuitization \& non-annuitization regions and can be solved for all $t < \tau$. This boundary can be used by the plan participant to develop an optimal strategy.  Initially, the plan participant will make sure that she stays at the annuitization boundary, ie., for a given wealth level, if she is inside the annuitization region, she will compute the amount $g_t$ that is needed to bring her to the annuitization boundary. No action is required if she is outside the annuitization region since DIAs cannot be shorted by the plan participant.  The plan participant re-evaluates her wealth, DIA levels, and repeats the process periodically.
\medbreak
Note that a verification theorem could be proved, showing that a sufficiently smooth solution to \eqref{post2} and \eqref{pre2} (including smooth pasting across the free boundary) with suitable finiteness conditions at $t=\infty$ and $w=\infty$, will yield a solution to our optimization problem. Our interest lies in numerical exploration of the solution's properties, and we will not carry out the required verification of smoothness and existence, therefore we do not claim a rigorous verification result here. 

\subsection{Solution methodology}
We solve the problem in two steps -- post-retirement followed by pre-retirement. We use implicit finite difference methods to solve the partial differential equation (\ref{post2}) and an explicit scheme to solve equation (\ref{pre2}).  We note that $\pi$ can be scaled to 1 without any loss of generality.  We also assume a CRRA choice of $U$. Boundary conditions are required for large $w$ and at $w=0$. Likewise, a terminal condition is required at large time for the post-retirement PDE (and at time $\tau$ for the pre-retirement one). 

\subsubsection{Boundary conditions:  post-retirement ($t\geq \tau$)}
At $t=\infty$, the Gompertz-Makeham \emph{Instantaneous Force of Mortality}, $\lambda_{x+t}$, increases exponentially (i.e. the client is dead) and asymptotically, from equation (\ref{post2}), we have 
\[
\lambda_{x+t} U(w+1)-\lambda_{x+t}J=0, 
\] 
which gives the terminal condition $J= U(w+1)$.

\medbreak
For maximum wealth in post retirement, we employ an asymptotic approximation at the boundary. We scale $w$ in equation (\ref{post2}) by a suitably large wealth value $w_m$ so that $W=w/w_m$ and $\delta=(1+I)/w_m\ll 1$.  After rescaling $J=w^{1-\gamma}_m\Phi(t,W)$ and $c^*_t=w_m C^*_t$, the HJB equation becomes
\begin{eqnarray}
0&=& \Phi_t+(\mu W+\delta-C^*_t)\Phi_W+\frac{1}{2}(\sigma W)^2\Phi_{WW} \\
&&+\lambda_{x+t}u(W+\delta)+u(C^*_t)-(\rho+\lambda_{x+t})\Phi \nonumber
\end{eqnarray}
where
\begin{equation}
C^*_t=\Phi^{-\frac{1}{\gamma}}_W.
\end{equation}

We now expand the value function as
\[
\Phi=\Phi^{(0)}(t,W)+\delta \Phi^{(1)}(t,W)+\cdots
\]
and the optimal consumption rate becomes $C^*_t=C^{(0)}_t+\delta C^{(1)}_t+\cdots$ where
\[
C^{(0)}_t=(\Phi^{(0)}_W)^{-\frac{1}{\gamma}},\; 
C^{(1)}_t=-\frac{1}{\gamma}(\Phi^{(0)}_W)^{-\frac{1}{\gamma}-1}\Phi^{(1)}_W.
\]

Substituting these expressions into the HJB equation, and collecting the zero'th and first order terms in $\delta$, we have
\begin{eqnarray}
0&=& \Phi^{(0)}_t+(\mu W-C^{(0)}_t)\Phi^{(0)}_W+\frac{1}{2}(\sigma W)^2\Phi^{(0)}_{WW} \\
&&+\lambda_{x+t}u(W)+u(C^{(0)}_t)-(\rho+\lambda_{x+t})\Phi^{(0)}, \nonumber \\
0&=& \Phi^{(1)}_t+(\mu W-C^{(0)}_t)\Phi^{(1)}_W+(1-C^{(1)}_t)\Phi^{(0)}_W+\frac{1}{2}(\sigma W)^2\Phi^{(1)}_{WW} \\
&&+u'(C^{(0)}_t)C^{(1)}_t+\lambda_{x+t}u'(W)-(\rho+\lambda_{x+t})\Phi^{(1)}, \nonumber
\end{eqnarray}
with terminal conditions $\Phi^{(0)}(T,W)=u(W)$ and $\Phi^{(1)}(T,W)=u'(W)$.

\medbreak
Using CRRA utility, we seek a solution in the form of $\Phi^{(0)}(t,W)=h(t)u(W)$ and $\Phi^{(1)}=k(t)u'(W)$, where $h(t)$ and $k(t)$ are the solutions of the following ODEs
\begin{equation}
\frac{dh}{dt}-\left[(\gamma-1)\mu+\rho+\lambda_{x+1}+\frac{\gamma(1-\gamma)\sigma^2}{2}\right] h+\gamma h^{\frac{\gamma-1}{\gamma}}+\lambda_{x+t}=0
\end{equation}
with $h(T)=1$ and

\begin{equation}
\frac{dk}{dt}-\left[\gamma\mu-\gamma h^{-\frac{1}{\gamma}}+\rho+\lambda_{x+1}-\frac{\sigma^2}{2}\gamma(1+\gamma)\right] k+\lambda_{x+t}+h=0
\end{equation}
with $k(T)=1$. Note that $\Phi^{(1)}>0$, the correction to the leading order approximation, is positive, which increases the value function.

\medbreak
For minimum wealth, we let $w\downarrow 0$. The two $w$ terms become much smaller than the rest of the equation (any potential singularity in $J_w$ or $J_{ww}$ being precluded, by virtue of model adding $\pi$ to the estate upon premature death), so the rest of the terms must balance themselves.
That is,
\begin{equation}
0= J_t+(I_t+1-c^*_t)J_w+\lambda_{x+t}  U(1) + U(c^*_t)-(\rho+\lambda_{x+t})J 
\end{equation}
where
$c^*=J^{-\frac{1}{\gamma}}_w.$

\subsubsection{Boundary conditions:  pre-retirement: $t<\tau$}
At $t=\tau$, the initial boundary is formed by the solution obtained for post-retirement.  We work over a domain  $0\leq I\leq I_{max}$, $0\leq w\leq w_{max}$ and $0\leq t\leq \tau$. Over a time-step we solve the PDE
\begin{equation}
\label{preretirementpde1}
0= J^{(1)}_t+(\mu w+\nu)J^{(1)}_w+\frac{1}{2}(\sigma w)^2J^{(1)}_{ww} 
+\lambda_{x+t}  U(w+K+\nu)-(\rho+\lambda_{x+t})J^{(1)}
\end{equation}
and
\begin{equation}
\label{preretirementpde2}
J^{(2)}_I - {}_{(\tau-t)}\tilde{a}_{(x+t)} J^{(2)}_w=0
\end{equation}
and then take $J=\max\{J^{(1)},J^{(2)}\}$.  The possible switching point is the \textbf{annuitization boundary} $w=w^*(I)$. 

\medbreak
Minimum wealth is always assumed to be in the non-annuitization region. As before, when $w\downarrow 0$, the two $w$ terms are much smaller than the rest of the equation, so the rest of the terms must balance themselves. In other words,
\begin{equation}
\label{preretirementminwealth}
0= J^{(1)}_t+(\nu)J^{(1)}_w
+\lambda_{x+t}  U({}_{(\tau-t)}\tilde{a}_{(x+t)} \cdot I_t+\nu)-(\rho+\lambda_{x+t})J^{(1)}.
\end{equation}

At maximum wealth, we perform a type of asymptotic analysis on the non-annuitization equation. We assume that $J(t,w,I) = U(w)V(t,I)$. With this assumption, we have 
\[J^{(1)}_w = \frac{(1-\gamma)}{w}U(w)V(t,I),\; \textrm{and}\;  J^{(1)}_{ww} = \frac{-\gamma(1-\gamma)}{w^2}U(w)V(t,I).\] Substituting this into equation (\ref{preretirementpde1}), and assuming that at maximum wealth $(\mu w+\nu) \approx \mu w$ and $(w+{}_{(\tau-t)}\bar{a}_{(x+t)} \cdot I_t+\nu) \approx w$, we obtain 
\begin{equation}
\label{preretirementpdemaxw1}
V_t+\left[\mu(1-\gamma)-\frac{1}{2}\gamma(1-\gamma)(\sigma)^2  -(\rho+\lambda_{x+t})\right]V+\lambda_{x+t}=0
\end{equation}
or 
\begin{equation}
\label{preretirementpdemaxw2}
J^{(2)}_I={}_{(\tau-t)}\tilde{a}_{(x+t)} J^{(2)}_w.
\end{equation}

What we do is solve for $V$ at each income level, and then multiply by $U(W_{max})$ to find the value of $J^{(1)}$. 

\subsection{Parameter values}

To generate our results we need capital market assumptions ($r, \mu, \sigma$), mortality parameters ($\lambda_0, m, b$), and a measure of risk aversion ($\gamma$). Our base-case capital market assumptions come from the 2017 Long-Term Capital
Market Assumptions from J.P. Morgan Asset Management. Specifically, we use the returns assumption of US Large Cap as a proxy for risky assets, i.e. $\mu = 7.25 \%$ and $\sigma = 14.75\%$, and US Short Duration Government Treasury as a proxy for risk-free return, i.e. $r = 3.25\%$. For simplicity, we round these values for implementation purposes, and take $\mu = 8 \%$ and $\sigma = 16\%$.  In the sensitivity analysis we adjust these parameters and compare results.  We do not adjust $\nu$ and $\pi$. We simply set them to 1 for all calculations. In other words, wealth is measured in multiples of the savings rate or the exogenous pension (which are assumed equal).  Our Gompertz-Makeham mortality model parameters are $\lambda_0 = 0.0$, $m = 89.335$, $b=9.5$. We take the subjective discount rate $\rho$ to be equal to the risk-free rate $r$. 

\medbreak
A variety of studies have estimated the value of $\gamma$. One of the earliest papers is the work by \cite{friend-blume-1975}, which has withstood the test of time and provides an empirical justification for constant relative risk aversion, estimates the value of $\gamma$ to be between 1 and 2.  \cite{feldstein-ranguelova-2001}; \cite{mitchel-poterba-warshawsky-brown-1999} in the economics literature have employed values of less than 3.  \cite{mankiw-zeldas-1991}; \cite{blake-burrows-2001}; \cite{campbell-viceira-2002} suggest that risk aversion levels might be higher. On the other hand, to avoid the problem of picking a $\gamma$ value, \cite{browne-milevsky-salisbury-2003} invert the Merton optimum to solve for $\gamma$.  However, any formulaic approach requires that we have the client’s complete financial balance sheet inclusive of financial and real assets. In this paper, based on our observations we chose a $\gamma$ value of 3 as the baseline value. 


\section{Results and discussion} \label{section:Observations1}
\begin{figure}[H]
  \captionsetup{font=small}
   \centering
   \includegraphics[width=0.5\textwidth, height=0.32\textheight]{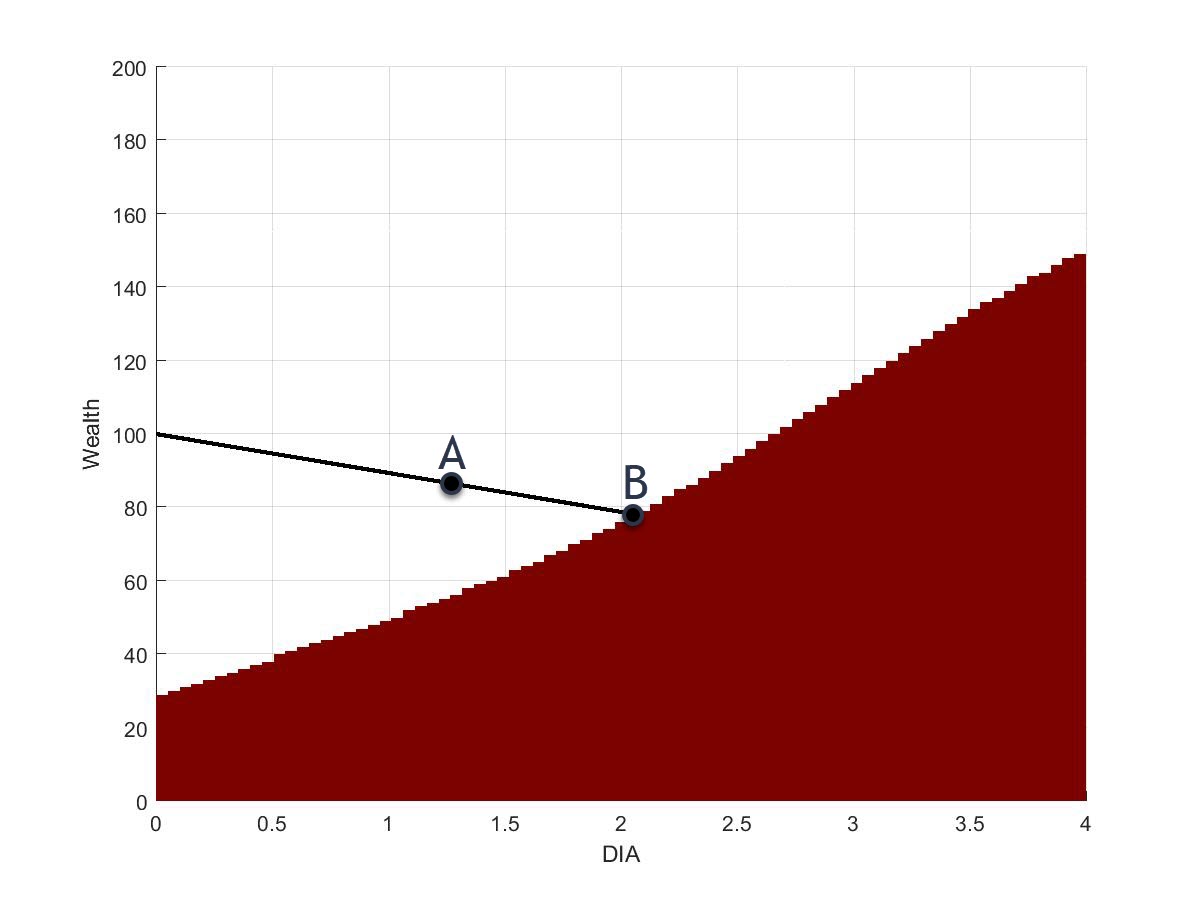}
   \caption{Investment profile at age 55 with the characteristic line included.  Units of DIA ($I$) are displayed on the $x$-axis while wealth levels ($W$) are on the $y$-axis.  For a given wealth and DIA, if the client is in the annuitization region (clear) then she will annuitize a portion of his wealth to arrive at the boundary between annuitization and non-annuitization.  On the other hand, if the client is in the non-annuitization region (dark) then no action will be taken as annuities cannot be sold to bring the client back to the boundary.  The slope of the characteristic line is $-_{\tau - t}\tilde{a}_{x+t}$. Baseline parameter values and $Q=1$.}\label{fig:charexamp}
\end{figure}

Figure~\ref{fig:charexamp} illustrates an optimal solution for an investor at age 55 when the DIA is fully refundable ($Q=1$). There are two regions on a Wealth-DIA ($I-w$) plane separated by an annuitization boundary, where annuitization is optimal inside the clear zone, while it is optimal not to annuitize in the darker region. The indexing on our axis is scaled by the pension term $\pi$. This means, for instance, that at the grid point with $W = 20$, and $I=1$, the investor has 20 times the value of their yearly exogenous pension in investment assets, and one times their yearly exogenous pension as a annuity income. The straight line marked by $A$ and $B$ is given by the equation 
\[\frac{dw}{dI}=-_{\tau - t}\tilde{a}_{x+t}\]
is the slope of the line for $x=55$. This is an annuitization line (or a characteristic line), which is a level curve of the value function by \eqref{boundary}, since on it we have 
\[
\frac{dJ}{dw}:=\frac{\partial J}{\partial w}+\frac{dI}{dw}\frac{\partial J}{\partial I}=0.
\]
Note that there are infinitely many (parallel) annuitization lines, since the return of the portfolio is stochastic; for an investor at age $A$ with $I$ units of DIA, the wealth $w$ may take any value. For any wealth-DIA combination ($I,w$) in the clear region, e.g. the point on the sample annuitization line given by the point labelled $A$, the optimal strategy is to annuitize, i.e., to move along the annuitization line to the point labelled $B$ on the annuitization boundary between the clear and dark zones.

\medbreak
It is worth elaborating on how a possible investor would use these graphs. Say the market aligns itself with the baseline parameter values that we use, and say the investor has a $\gamma$ value of 3. At any point in time, the investor would look at her current wealth and income levels, and if she found herself in an annuitization region, she would annuitize until she reaches the boundary between the two regions. Of course, the cost of the annuity at time $t$ would be $_{\tau - t}\tilde{a}_{x+t}$. So for every small amount of DIA ($\delta I$) that the investor purchases, it will cost $\delta w=_{\tau - t}\tilde{a}_{x+t}\delta I$ liquid wealth. This is indicated by the straight line from any point, e.g., $A$, in the annuitization region to the point $B$ on annuitization boundary with a slope of $-_{\tau - t}\tilde{a}_{x+t}$. 
 
\medbreak

\begin{figure}[H]
  \captionsetup{font=small}
  \centering
  \begin{subfigure}[b]{0.48\textwidth}
  \includegraphics[width=\linewidth, height=0.40\textheight]{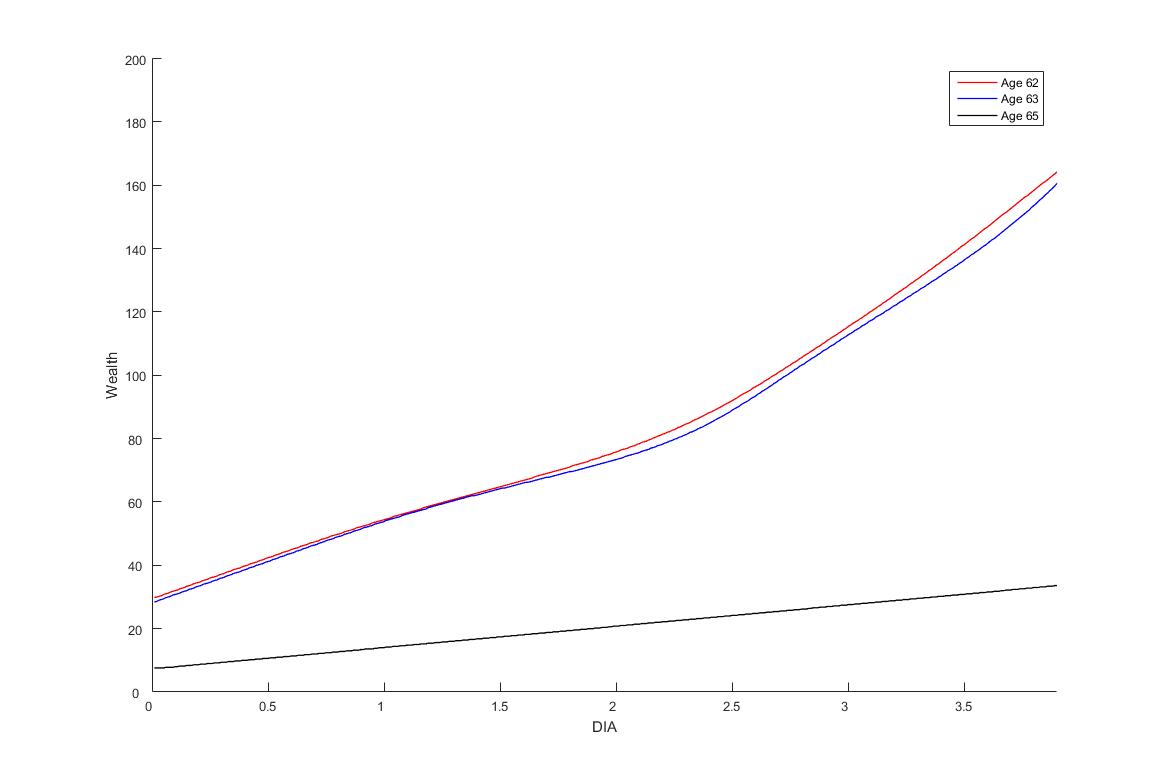}
  \caption{Investment Profile (at age 62, 63, and 65) for baseline parameter values and $Q=1$. }
  \label{fig:default}
  \end{subfigure}\hskip.2in
  \begin{subfigure}[b]{0.48\textwidth}
    \includegraphics[width=\linewidth, height=0.4\textheight]{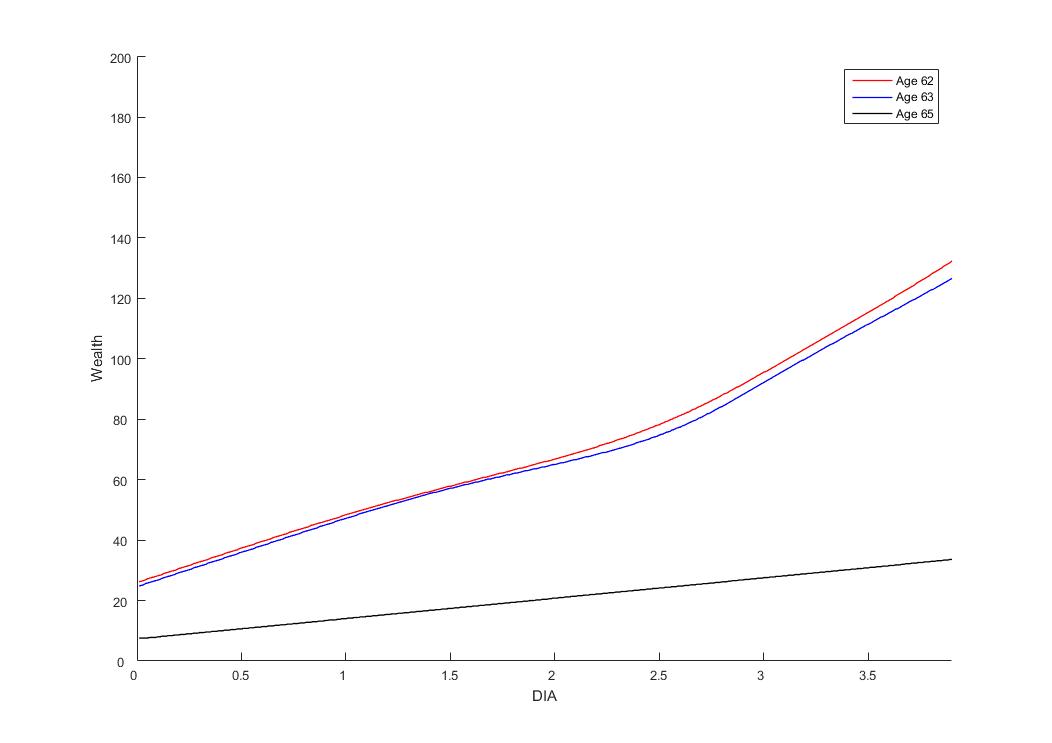}
  \caption{Investment profiles (at age 62, 63, and 65) for baseline parameter values and $Q=0.7$. }
  \label{fig:lqdefault}
 \end{subfigure}
    \caption{Investment profile for various ages with baseline parameter values for fully and partially refunded DIAs. The region above the curve is the annuitization region.}
    \label{FIG:AgeGraphBase}
\end{figure}

Figure \ref{FIG:AgeGraphBase} shows the annuitization boundary for the baseline parameter values ($\mu$, $\sigma$, $\gamma$) with $Q=1$ and $Q=0.7$. This means that in the former case there are no mortality credits. In the latter case, $70\%$ of the DIA is made up of a refundable DIA (with no moratlity credits) while $30\%$ is made up on a non-refundable DIA (with mortality credits). The annuitization region expands when time moves forward from age 62 to 65.  At earlier ages, we see a smaller annuitization region compared to the one at retirement.  The results are consistent with the standard idea that as a person gets older, he/she moves away from risky assets and invests more in riskless ones (in this case, the DIA acts as the riskless asset).  When we compare Figure \ref{fig:default} and Figure \ref{fig:lqdefault}, we can see that a lower $Q$ value has a larger annuitization region.  Since a lower $Q$ value creates mortality credits in the DIA, it therefore makes the DIA a more desirable product, from an income perspective. 

Notice the significant change in the annuitization region which takes place in the final time step of pre-retirement  (black lines in Figure~\ref{FIG:AgeGraphBase}).  Since our model does not permit the  investor to purchase additional annuities in retirement, the final time-step is the last opportunity to do so.  Furthermore, as the investor enters into retirement, she now has to think about consumption. So where as before she did not need to worry about having a steady income, and simply having wealth was enough, she now must prioritize consumption, and therefore annuitization is more attractive at the last time step.  Essentially, both her options available and her behaviour exhibit an abrupt change at retirement, as shown in the graphs.

\medbreak
The annuitization boundary moves continuously in time prior to retirement, which means that DIA will be purchased continuously, with the possible exception of lump-sum purchases at initiation and at retirement. If market noise (or movement of the boundary) acts to move wealth up into the annuitization region, DIA purchase will act to keep $(I,w)$ at the boundary. Whereas noise can move $(I,w)$ below the boundary, in which case DIA purchase ceases, at least temporarily. 

\begin{figure}[H]
  \captionsetup{font=small}
  \centering
  \begin{subfigure}[b]{0.48\textwidth}
   \includegraphics[width=\linewidth, height=0.40\textheight]{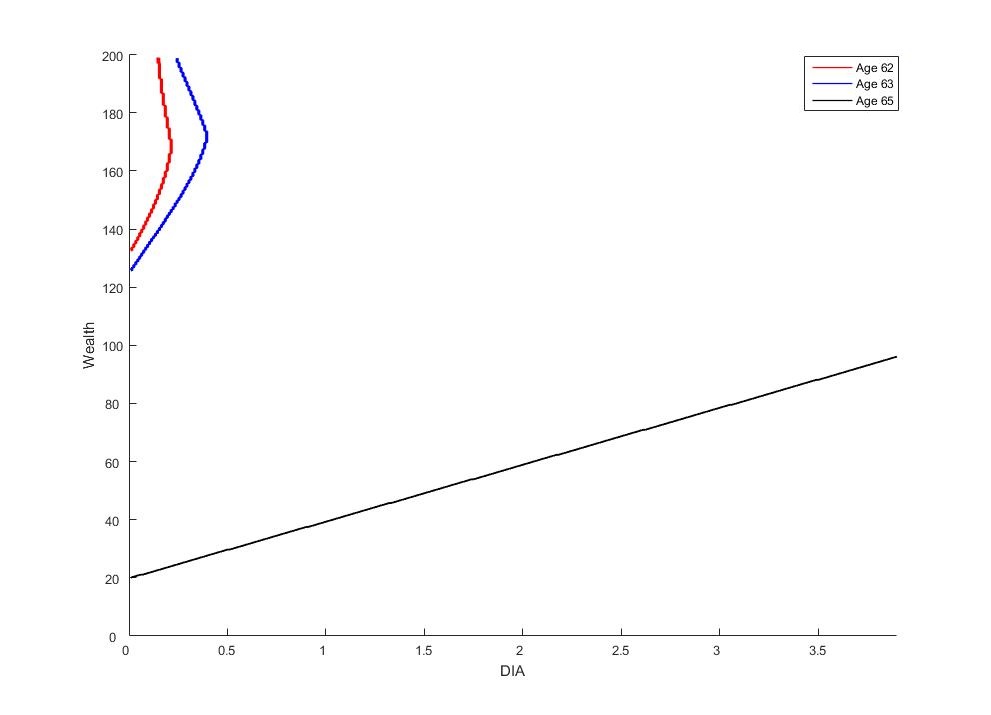}
   \caption{Investment Profile (at age 62, 63, and 65) for $\mu=10\%$ and $Q=1$.  }
   \label{fig:HiMu}
  \end{subfigure}\hskip.2in
  \begin{subfigure}[b]{0.48\textwidth}
    \includegraphics[width=\linewidth, height=0.4\textheight]{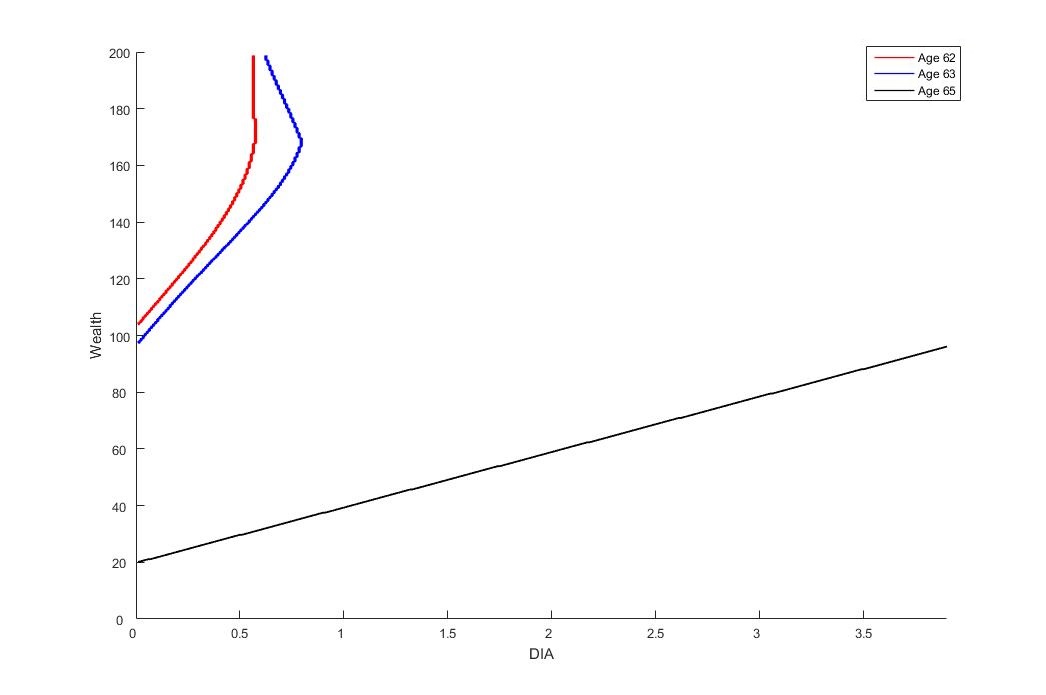}
  \caption{Investment profiles (at age 62, 63, and 65) for $\mu=10\%$ and $Q=0.7$.  }
   \label{fig:lqHiMu}
  \end{subfigure}
    \caption{Investment profile for various ages with higher $\mu$ for fully and partially refunded DIAs. Other parameters remain unchanged. The annuitization region lies above the curves.}
    \label{FIG:AgeGraphHighMu}
\end{figure}

\medbreak
Next, we vary the rate of return of the risky asset ($\mu$) to investigate its impact on annuity allocation. The region below the three curves (for three different ages) is the non-annuitization region while the region above the curves is the annuitization region. As can be seen from Figure~\ref{FIG:AgeGraphHighMu}, the change in the annuitization boundary is significant. With larger $\mu$ values, equity investments are more attractive, so the annuitization region is smaller, as expected. The annuitization boundary is clearly not monotonic -- at some ages and levels of $I$, it is never optimal to purchase additional DIAs, regardless of $w$.

\medbreak
We do not have a good explanation for the apparent kink observed in Figure \ref{FIG:AgeGraphBase} (and some other figures) between $I=2$ and $I=3$. But we have performed a variety of sensitivity checks, to ensure that it is not an artifact of our numerical scheme. There is a related feature in Figure \ref{FIG:AgeGraphHighMu}, namely that the annuitization boundary may not be monotonic. In fact, if we had shrunk the scale to show values of wealth that are implausibly high economically, it would be apparent that our mathematical problem can have the annuitization boundary curl back and cross the vertical axis. This must be due to either labour or pension income, because otherwise scale invariance would apply, and any annuitization boundary would be linear. In fact, numerical experiments suggest that it is pension income that drives this feature.

\medbreak
To understand this, observe that there is partial scale invariance in our problem, in the sense that increasing wealth (with pension income fixed) is equivalent to decreasing pension income (with wealth held fixed). With sufficiently large pension income, we of course don't annuitize at all, since we are effectively already overweighted with bond-like assets. So the feature of interest corresponds to a situation where with no pension income we would annuitize only at retirement, but with moderate pension income, some DIA purchase becomes optimal.

\medbreak
With no pension income, at retirement we would optimally annuitize a fixed fraction of wealth. But once pension income becomes significant, that will not be the case -- a massive collapse in equity prices would cause us to forego annuitization at retirement, and instead rely on pension income for the bond-like piece of our post-retirement portfolio. We conjecture that this causes the effect we observe, ie. that fear of missing out on annuitization at retirement may cause us to lock in some annuity right away. This effect deserves further study.

\begin{figure}[H]
\captionsetup{font=small}
  \centering
    \begin{subfigure}[b]{0.48\textwidth}
      \includegraphics[width=\linewidth, height=0.25\textheight]{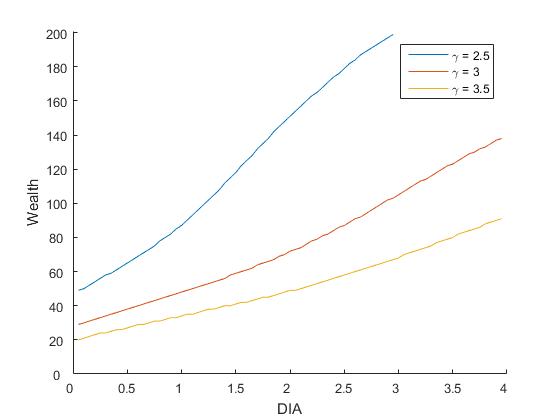}
      \caption{Multiple regime boundaries at age 55.}
      \label{FIG:MultiGammaAge55}
    \end{subfigure}
    \quad
    \begin{subfigure}[b]{0.48\textwidth}
      \includegraphics[width=\linewidth, height=0.25\textheight]{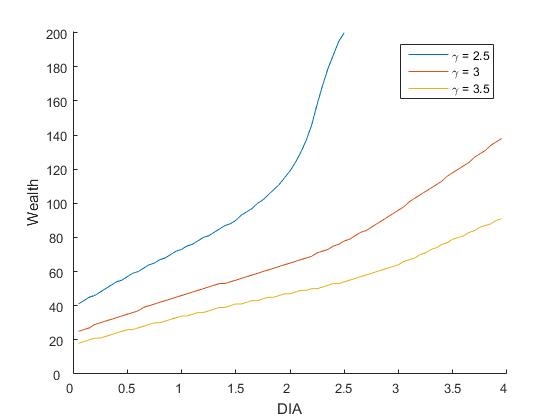}
      \caption{Multiple regime boundaries at age 61.}
      \label{FIG:MultiGammaAge61}
    \end{subfigure}
    \caption{Investment Profile for various values of $\gamma$ (Coefficient of Risk Aversion) using baseline parameters and $Q=1$}
    \label{FIG:GammaGraph}
\end{figure}

\medbreak
When we increase the value of the risk-aversion coefficient ($\gamma$), the annuitization region gets larger as can been seen in Figure \ref{FIG:GammaGraph}.  Again, the region below the curve is the non-annuitization region while we annuitize when above the curve.  A more risk-averse individual (higher $\gamma$) purchases more DIA, as we would expect.

\medbreak
\section{Optimal asset and DIA allocation} \label{section:AssetAllocation}
The analysis in the preceding section kept asset allocation constant, i.e. the plan participant does not in any way adjust the allocation to risky asset.  
In this section, we introduce choices for asset allocation by allowing the plan participant to invest in an optimal combination of risky and riskless assets. We compute at each time step the optimal mix of risky and riskless assets, and assume that plan participants will invest in this optimal mix. In contrast, our earlier analysis had GBM dynamics for the investment account, which corresponds to a fixed (exogenous) asset allocation. 

\medbreak
We postulate a riskless asset for which the rate of return is $r$.  We use a variable $\alpha$ which ranges from 0 to 1 to specify what proportion of our investment portfolio is allocated to the risky asset.  The rate of return and volatility of this new portfolio are  $\alpha \mu + (1-\alpha)r$ and $\alpha \sigma$, respectively.

\medbreak
To reformulate our equations, we simply replace every $\mu$ and $\sigma$ term in the previous section, with these new expressions for rate of return and volatility.  Our pre-retirement wealth process is
\begin{equation}
dW_t = ((\alpha_t \mu + (1-\alpha_t)r)W_t + \nu -g_t)dt + \sigma \alpha_t W_t dB_t
\end{equation}

And our post-retirement wealth process is
\begin{equation}
dW_t = ((\alpha_t \mu + (1-\alpha_t)r)W_t + I_t +\pi -c_t)dt + \sigma \alpha_t W_t dB_t.
\end{equation}
Our value function $J$ is defined as before (except that $\alpha_t$ is also a control variable), and so we can apply the same Bellman optimality
principle to find
\begin{eqnarray}
0&\geq& J_t+((\alpha_t \mu + (1-\alpha_t)r) w+\nu-g_t)J_w+\frac{1}{2}(\sigma \alpha_t w)^2J_{ww} \\
&&+\frac{g_t}{{}_{(\tau-t)}\tilde{a}_{(x+t)}}J_I+g_tJ_K+\lambda_{x+t} U(w+K+\nu)-(\rho+\lambda_{x+t})J \nonumber
\end{eqnarray}
for $t<\tau$, and
\begin{eqnarray}
0&\geq& J_t+((\alpha_t \mu + (1-\alpha_t)r) w+I_t+\pi-c_t)J_w+\frac{1}{2}(\sigma \alpha_t w)^2J_{ww} \label{post}\\
&&+\lambda_{x+t}  U(w+\pi)+U(c_t)-(\rho+\lambda_{x+t})J \nonumber
\end{eqnarray}
for $t\geq\tau$.

\medbreak
We follow the exact same procedure as before, as in Equations (\ref{firststep}) through (\ref{laststep}). The only difference is that we need to find the optimal asset allocation $\alpha_t$. This is done by applying the first-order condition, i.e., taking the first derivative with respect to $\alpha_t$ in both equations and setting it equal to zero. Solving for alpha we get
 $$\alpha^*_t = -\frac{J_{w}}{J_{ww}}\frac{\mu - r}{w \sigma ^2}.$$ 

\medbreak
In post-retirement, we solve
\begin{eqnarray}
\label{postretirementpdealpha}
0&=& J_t+((\alpha^*_t \mu + (1-\alpha^*_t)r) w+I_t+1-c^*_t)J_w+\frac{1}{2}(\sigma \alpha^*_t w)^2J_{ww} \\
&&+\lambda_{x+t} U(w+1)+U(c^*_t)-(\rho+\lambda_{x+t})J \nonumber
\end{eqnarray}
and in pre-retirment, we solve the PDE
\begin{equation}
\label{preretirementpde1alpha}
0= J^{(1)}_t+((\alpha^*_t \mu + (1-\alpha^*_t)r) w+\nu)J^{(1)}_w+\frac{1}{2}(\sigma \alpha^*_t w)^2J^{(1)}_{ww} 
+\lambda_{x+t} U(w+K+\nu)-(\rho+\lambda_{x+t})J^{(1)}
\end{equation}
below the annuitization boundary, and 
\begin{equation}
\label{preretirementpde2alpha}
J^{(2)}_I+{}_{(\tau-t)}\tilde{a}_{(x+t)} \cdot (J^{(2)}_K-J^{(2)}_w)=0
\end{equation}
above it.

\medbreak
The method we employ to solve these new equations is precisely the same as the previous section. We reformulate the finite difference equation with new coefficients as stated above and adjust the terminal and boundary conditions.

\medbreak
Mathematically there is nothing stopping us from having an allocation to the risky asset which is greater than $100 \%$,  but most retirement portfolios are subject to a restriction that one cannot borrow money and leverage the portfolio. Likewise, a short position (i.e. $\alpha_t < 0$) within retirement accounts is also normally prohibited.  Therefore, we constrain $0 \leq \alpha^*_t \leq 1$ when calculating the optimal $\alpha^*_t$.

\subsection{Boundary conditions}
 As discussed previously, in post-retirement, we perform a very similar asymptotic approximation to get our boundary condition at maximum wealth.  The process is the same except with adjusted drift and volatility terms.  At the maximum wealth level, the effects of income from the DIA are negligible so we assume that the problem reverts to the standard Merton portfolio problem and we use the $\alpha^*$ value from the standard Merton problem, i.e.  
 $$\alpha^* = \frac{(\mu - r)}{\gamma \sigma^2}.$$
 For further details on the derivation of this equation, see \cite{Rogers2013}.  Our boundary conditions remain unchanged in pre-retirement.

\subsection{Results and discussion} \label{section:Observations2}
\subsubsection{Asset allocation only in pre-retirement}
We are interested is seeing how the decision to purchase a DIA will be affected if asset allocation is allowed.  To start with, we keep the asset allocation fixed in post-retirement (as in the previous section) so as not to change the post-retirement consumption profile.  This allows a direct comparison with the results of the previous section, and in Section \ref{dynamicallocationpostretirement} we will go on to allow dynamic asset allocation in both pre- and post-retirement. 

\medbreak
When reading the graphs in this section, note that the part of the graph which is clear is the annuitization region (i.e. above the annuitization boundary), while the colour-banded region gives the asset allocation ($\alpha^*$) profile. The color bar indicates what percentage of the portfolio is made up of the risky asset.  For comparison, the outline of the annuitization boundary from the fixed asset allocation model is also displayed (black line). 

\begin{figure}[H]
\captionsetup{font=small}
    \centering
    \begin{subfigure}[b]{0.45\textwidth}
    \includegraphics[width=\linewidth, height=0.32\textheight]{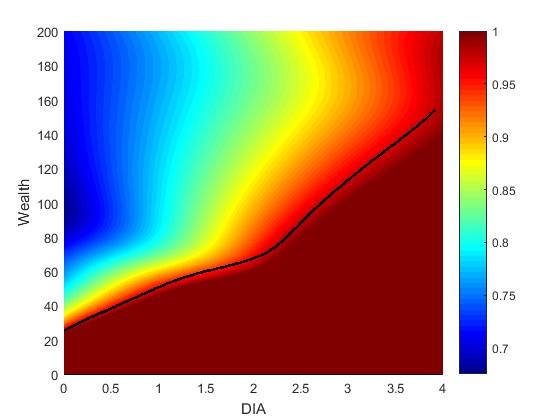}
    \caption{Investment profile at age 64.}
       \label{fig:boundrycompare64}
    \end{subfigure}
    \quad
    \begin{subfigure}[b]{0.45\textwidth}
      \includegraphics[width=\linewidth, height=0.32\textheight]{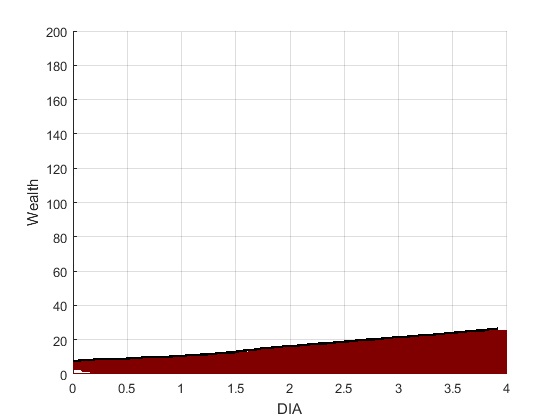}
       \caption{Investment profile at age 65.}
      \label{fig:boundrycompare65}
    \end{subfigure}
    \caption{Investment profile with optimal asset allocation in pre-retirement and fixed asset allocation in post-retirement (baseline parameters and $Q=1$).  The color band is an indication of the allocation to risky assets.  The black line shows the result from the previous section where asset allocation is not available.  The clear region is the annuitization region.}
\end{figure}

We start with baseline parameters and compare with the results from the previous section (Figures \ref{FIG:AgeGraphBase} through \ref{FIG:GammaGraph}).
When $Q = 1$, the DIA carries no mortality credits, and therefore indistinguishable from the risk free asset, except that the purchase of a DIA is irreversible.  Therefore, an investor who wishes to purchase a DIA should prefer to invest risk-free and wait until retirement to purchase a no delay DIA (effectively, a SPIA). That way they avoid the possibility that equity movements and the liquidity constraint lead to the DIA allocation at retirement being suboptimal. And indeed, there is no clear (annuitization) region in Figure \ref{fig:boundrycompare64}. 

\medbreak
Note how closely the old  annuitization boundary matches the region where it is optimal to allocate 100\% to the risky asset. For $(I,w)$ along the black curve in Figure \ref{fig:boundrycompare64}, the DIA is a proxy for the risk-free asset, so it is optimal to hold almost none of the latter.  At the final time step (Figure \ref{fig:boundrycompare65}), the optimal level of DIA purchase matches what we had earlier (as it should, since the post-retirement problems are identical). 

\begin{figure}[H]
  \captionsetup{font=small}
    \centering
    \includegraphics[width=0.5\textwidth, height=0.28\textheight]{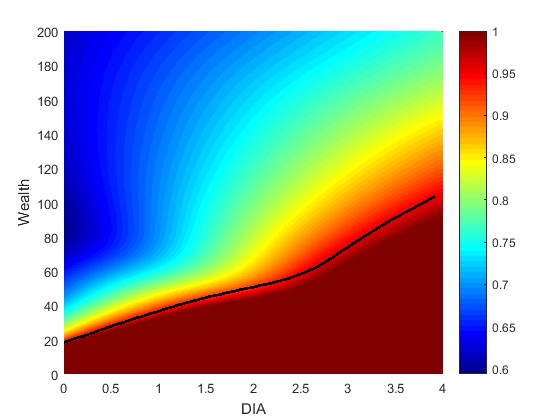}
       \caption{Investment profile with optimal asset allocation in pre-retirement and fixed asset allocation in post-retirement for $\gamma$ = 3.5 (other parameters at baseline)  at age 63 and $Q=1$.  The color band is an indication of the allocation to risky assets.  The black line shows the result from the previous section where asset allocation is not available.}
        \label{fig:boundrycompare63gamma3.5}
\end{figure}

\medbreak
In Figure \ref{fig:boundrycompare63gamma3.5} we show the impact of increased $\gamma$ (i.e. the client is relatively more risk averse). As expected, we still observe no purchasing of DIA before retirement.  What changes is that at all values of $(I,w)$, the client allocates more to the risk-free asset than before.

Next we present results for $Q = 0.7$, which means that mortality credits are present, and therefore we are incentivizing the early purchase of the DIA. We now see that there are situations where both the risk free asset and the DIA are viable in a time step, depending on ones wealth and income level. We begin with baseline parameters. 

\begin{figure}[H]
\captionsetup{font=small}
  \centering
  \begin{subfigure}[b]{0.45\textwidth}
    \includegraphics[width=\linewidth, height=0.28\textheight]{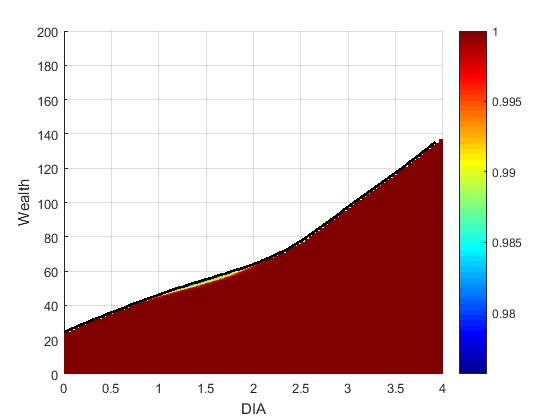}
    \caption{Investment profile at age 62}
    \label{fig:lqboundrycompare62}
  \end{subfigure}
  \quad
  \begin{subfigure}[b]{0.45\textwidth}
     \includegraphics[width=\linewidth, height=0.28\textheight]{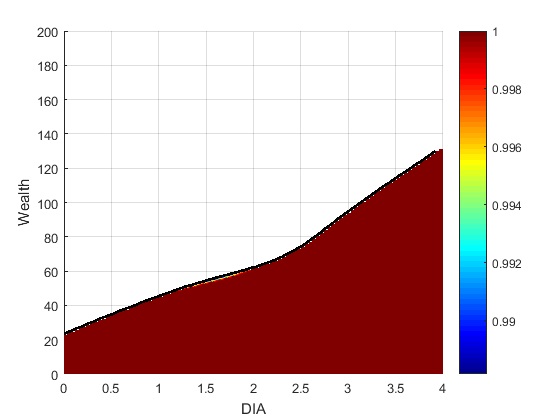}
     \caption{Investment profile at age 63}
     \label{fig:lqboundrycompare63}
  \end{subfigure}
  \caption{Investment profile with optimal asset allocation in pre-retirement and fixed asset allocation in post-retirement (baseline parameters) for $Q=0.7$;  The color band is an indication of the allocation to risky assets.  The black line is the result from the previous section where asset allocation is not available.}
\end{figure}

\begin{figure}[H]
\captionsetup{font=small}
  \centering
  \begin{subfigure}[b]{0.45\textwidth}
    \includegraphics[width=\linewidth, height=0.25\textheight]{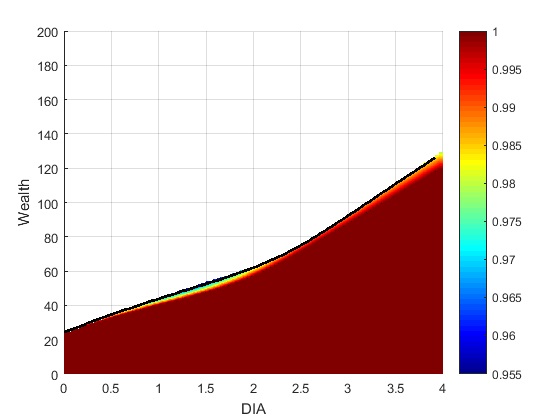}
    \caption{Investment profile at age 60}
    \label{FIG:ALPHA_G31_Q07_SIGMA16_MU08_AGE60}
  \end{subfigure}
  \quad
  \begin{subfigure}[b]{0.45\textwidth}
    \includegraphics[width=\linewidth, height=0.25\textheight]{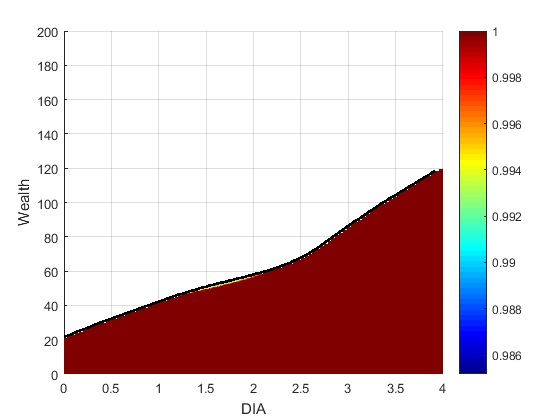}
    \caption{Investment profile at age 63}
    \label{FIG:ALPHA_G31_Q07_SIGMA16_MU08_AGE63}
  \end{subfigure}
  \caption{Investment profile with optimal asset allocation in pre-retirement and fixed asset allocation in post-retirement (baseline parameters with $\gamma=3.1$) for $Q=0.7$;  The color band is an indication of the allocation to risky assets.  The black line is the result from the previous section where asset allocation is not available.}
  \label{FIG:ALPHA_G31_Q07_SIGMA16_MU08}
\end{figure}

We see at age 62 (Figure \ref{fig:lqboundrycompare62}) that there is a large annuitization region, and also a large region where 100\% of the investment account is invested in the risky asset. In between there is only a small region in which the investor invests does allocate to the risk free asset. In this model, the DIA is preferred to the risk-free asset, except at times where one has just purchased DIA (where the liquidity of the risk-free asset allows one to adjust the allocation to the risky asset up, if its price declines). At a later time (Figure \ref{fig:lqboundrycompare63}), we see that the region of risk free investing becomes all but non-existent.  As the investor gets older, she moves away from the risk free investment and instead uses the DIA in its place.  With an increase in risk aversion, we observe a similar co-mingling of DIA, risk-free asset, and risky asset (Figure \ref{FIG:ALPHA_G31_Q07_SIGMA16_MU08}).

\subsubsection{Optimal allocation in post retirement}
\label{dynamicallocationpostretirement}
Finally, we present some results in which asset allocation is also allowed post retirement.  We first show the results for $Q=1$ and then for $Q=0.7$.   We start with baseline parameters and compare them with previous results.

\begin{figure}[H]
\captionsetup{font=small}
  \centering
  \begin{subfigure}[b]{0.45\textwidth}
    \includegraphics[width=\linewidth, height=0.25\textheight]{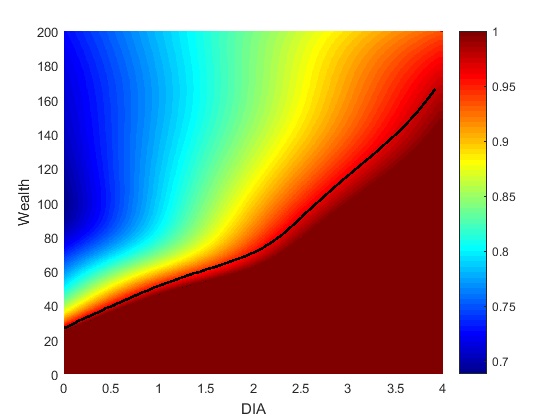}
    \caption{Investment profile at age 63}
    \label{fig:postalphaboundrycompare63}
  \end{subfigure}
  \quad
  \begin{subfigure}[b]{0.45\textwidth}
   \includegraphics[width=\linewidth, height=0.25\textheight]{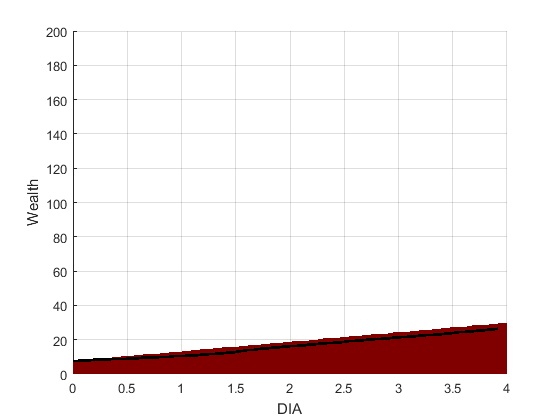}
   \caption{Investment profile at age 65}
   \label{fig:postalphaboundrycompare65}
 \end{subfigure}
 \caption{Investment profile with optimal asset allocation in pre-retirement and post-retirement (baseline parameters) for $Q=1$.  The color band is an indication of the allocation to risky assets.  The black line is the result from the previous section where asset allocation is not available.}
  \label{FIG:ALPHA_ALL_G3_Q10_SIGMA16_MU08}
\end{figure}

\begin{figure}[H]
\captionsetup{font=small}
   \centering
    \includegraphics[width=0.5\textwidth, height=0.32\textheight]{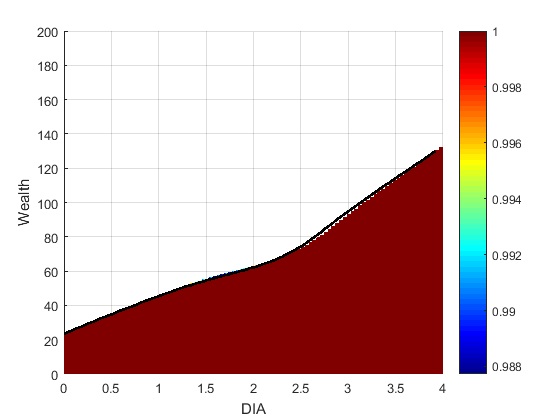}
       \caption{Optimal asset allocation (in pre and post retirement) and fixed asset allocation at age 63, with $Q=0.7$ (baseline parameters).}
        \label{fig:lqpostalphaboundrycompare63}
\end{figure}

As before, at $Q = 1$ (Figure \ref{FIG:ALPHA_ALL_G3_Q10_SIGMA16_MU08}), the DIA has no mortality credits, and therefore it is indistinguishable from the risk free asset, except for the non reversibility of the DIA.  So there is no DIA purchase prior to retirement. The annuitization region at retirement no longer perfectly matches the earlier versions, but the difference is small. Likewise with $Q=0.7$, we see only a small difference in strategy when compared to the fixed allocation results -- see Figure \ref{fig:lqpostalphaboundrycompare63}. The reason post-retirement asset allocation has so little impact may be that once consumption starts, wealth declines, and moves one to the region where it is optimal to allocate only to the DIA and the risky asset.

\section{Conclusion}
In this paper we explored how one should optimally purchase DIAs within a retirement investment plan.  We developed an optimal allocation strategy using the principles of stochastic control theory and dynamic programming.  This gives the wealth \& income levels at which it would be optimal to annuitize a portion of the investments into a DIA in pre-retirement. 

\medbreak
We considered DIA purchase under both a fixed and a dynamic asset allocation strategy.  While the DIA allocation region is larger under the fixed allocation strategy it is so because the DIA allocation becomes a proxy to fixed-income allocation.  Under dynamic asset allocation, the allocation to a refundable DIA is less appealing in comparison to a DIA carrying mortality credits.  

\medbreak
Our model implicitly assumes that mortality rates are predictable and that the hazard rate is deterministic.  There is a growing body of research into stochastic mortality models which it would be interesting to incorporate into our model.  Since DIA prices are driven by the combined effects of mortality and prevailing interest rates, incorporating interest rate volatility into our model would also be of interest. We leave these as questions for future research.

\medbreak
Finally, our model shows that periodic purchases of DIAs well before retirement can be attractive, but only if the refund provisions allow the products to retain significant mortality credits. 
    
\medbreak
\subsection*{Acknowledgement}
Mauskopf's research in this paper was partly supported by the Mitacs Accelerate program.  Huang and Salisbury's research is supported in part by NSERC and the Fields Institute. CANNEX Financial Exchanges has a commercial interest in the work presented in this paper.

\clearpage
\nocite{jpm-ltcma-2018}
\nocite{soa-rp2014}
\bibliographystyle{jf}
\bibliography{References}

\end{document}